\documentclass[twocolumn,nofootinbib,showpacs,prd]{revtex4-1}%
\usepackage{amsmath}
\usepackage{amsfonts}
\usepackage{amssymb}
\usepackage{graphicx}
\usepackage{color}

\begin{document}

\title{Modeling dark energy through an Ising fluid with network interactions}

\author{Orlando Luongo}
\affiliation{Physics Department, University of Naples "Federico II", I-80126, V. Cinthia, Naples, Italy,}
\affiliation{INFN, Section of Naples, I-80126, Naples, Italy}
\affiliation{Institute of Nuclear Sciences, UNAM, AP 70543, Mexico, DF 04510, Mexico.}

\author{Damiano Tommasini}
\affiliation{Institute of Physics and MTA-DE Particle Physics Research Group, University of Debrecen,
H-4010, Debrecen, P.O. Box 105, Hungary.}

\begin{abstract}
We show that the dark energy effects can be modeled by using an \emph{Ising perfect fluid} with network interactions, whose low redshift equation of state, i.e. $\omega_0$, becomes $\omega_0=-1$ as in the $\Lambda$CDM model. In our picture, dark energy is characterized by a barotropic fluid on a lattice in the equilibrium configuration. Thus, mimicking the spin interaction by replacing the spin variable with an occupational number, the pressure naturally becomes negative. We find that the corresponding equation of state mimics the effects of a variable dark energy term, whose limiting case reduces to the cosmological constant $\Lambda$. This permits us to avoid the introduction of a vacuum energy as dark energy source by hand, alleviating the coincidence and fine tuning problems. We find fairly good cosmological constraints, by performing three tests with supernovae Ia, baryonic acoustic oscillation and cosmic microwave background measurements. Finally, we perform the AIC and BIC selection criteria, showing that our model is statistically favored with respect to the CPL parametrization.
\end{abstract}


\maketitle

\section{Introduction}
\label{introduzione}

Recent cosmological observations pointed out that the universe is undergoing an accelerated expansion \cite{Riess:1998cb,Perlmutter:1998np}.
Unfortunately, the physical mechanism which drives the observed cosmic speed up is so far unclear. Moreover, standard pressureless baryonic matter is inadequate by itself to characterize the universe acceleration, even by assuming the additional presence of cold dark matter \cite{de Bernardis:2000gy,Percival:2002gq,Croft:1997jf,McDonald:2004eu}. As a standard landscape, it is possible to postulate the existence of a further ingredient, dubbed  dark energy (DE) \cite{Sahni:2006pa,Clarkson:2011zq,Kolb:2004am,Rasanen:2006kp,bll}. Dark energy behaves as a weakly interacting anti-gravitational fluid, described by a negative equation of state (EoS). Even though the nature of such a fluid has  not been clarified, a wide number of different paradigms have followed each other, spanning from slowly rolling scalar field, known as quintessence \cite{Ratra:1987rm,mioarticolo}, lattice of topological defeats
\cite{Vilenkin}, to barotropic fluids \cite{lindox} or modifications of Einstein gravity \cite{Capozziello:2003tk,Carroll:2003wy,Carroll:2004de} and so forth (for further details see \cite{Copeland:2006wr} and references therein). One of the simplest way to explain the observed cosmic speed up is provided by the so-called $\Lambda$CDM paradigm. In particular, the $\Lambda$CDM model leads to the introduction of a vacuum energy cosmological constant $\Lambda$, and assumes a total matter content, $\Omega_m$, given by the sum of baryonic and dark matter densities \cite{Carroll:1991mt,Tegmark:2003ud}. The corresponding EoS is constant as the universe expands, leading moreover to a negative and constant pressure \cite{glock}. Even though the model is in a fairly good agreement with current observations, it suffers from two profound shortcomings, i.e. the fine tuning and coincidence problems  \cite{Copeland:2006wr,Ratra:1987rm,Caldwell:1997ii,pad}. To alleviate these two issues, the DE density, $\rho$, and its corresponding pressure, $\mathcal{P}(\rho)$, are thought to evolve separately with significative departures from standard matter. Thus, possible extensions of the $\Lambda$CDM model are frequently characterized by assuming a time-variable EoS, $\omega(z)\equiv\frac{\mathcal{P}(\rho)}{\rho}$, evolving as $-1\leq\omega<0$, at $z=0$.

In this work, we investigate how to model DE by considering an Ising network-interacting fluid on a lattice \cite{mioarticolo,P88,Prigogine:1989zz,W}. We show that a fluid, whose interaction is provided by a series of networks, may predict effects due to a negative pressure, at $z\ll1$ \cite{qux}. Thence, the DE nature may be approximated by a network interacting Ising fluid, simply postulating the validity of standard thermodynamics, in the equilibrium configuration. In particular, one can assume that DE is modeled in analogy to the case of Ising chains. In our model, the role played by spin variables is replaced by occupational numbers. In doing so, one mimics the spin interaction without the need of real spin chains. The main advantage is to recover the physical properties of a spin interacting fluid in cosmology, predicting repulsive effects, able to explain the observed late time DE. The procedure of mimicking an Ising system has been extensively investigated in the literature. Examples have been considered in the case of liquid-gas phase transitions for atomic systems \cite{Lee}.

This paper is structured as follows. In Sec. II, we show the main features of our model. We focus on its thermodynamic interpretation, with particular attention to relate these results with the Friedmann equations. In Sec. III, we derive the cosmological model and the corresponding EoS. We investigate the acceleration parameter and its variation. We find out theoretical limits on the observable quantities and we determine the transition redshift, at which the acceleration starts. In Sec. IV, we perform three experimental tests, with supernovae Ia (SNeIa), baryonic acoustic oscillation (BAO) and cosmic microwave background (CMB), in order to constrain our model. In Sec. V, we adopt the AIC and BIC selection criteria, for comparing our model with alternative approaches. Finally, Sec. VI is devoted to conclusion and perspectives of our work.


\section{The Ising fluid with network interactions}
\label{secondasezione}

In this section, we describe the cosmological consequences of assuming an Ising fluid with network interactions, as a source of DE. Let us first notice that in  standard lattice models, the particle description is formally represented by assuming a grid with an occupational variable $\sigma_i$. Its value is zero, if the site is empty, and one if occupied by particles. This simple picture is also known in the literature as "bit`` gas model, in analogy to computational science \cite{kiki,kiki2}. Its use is usually adopted in different fields of physics, especially in order to model extended sites by numerical computations through Monte Carlo simulations, ranging from condensed matter to quantum computing and particle physics \cite{Lehaut:2008ix,Wilding,Lu:2008zzg,Souza,Simon,Higuera,Ruelle,Qian,Dieter}. Under these hypotheses, the occupational variables $\sigma_i$ are actually analogous to spin variables in the well known Ising model \cite{Kennedy,Binder}. However, the physical meaning of $\sigma_i$ is basically different from Ising spins. In particular, $\sigma_i$ is not an intrinsic physical property of the system, but only a way to discriminate the presence and absence of particles. Moreover, once the particle interaction is specified, our lattice fluid may lead to a negative EoS, in particular regions of the phase space \cite{b11,b12}. We will consider this property, in order to describe the DE effects in a homogeneous and isotropic universe.

\subsection{The entropy representation}

In the Hamiltonian formalism, we can write down the Hamiltonian of our model, by defining a chemical potential $\mu$. We have
\begin{equation}\label{ham}
{\mathcal H}=-\sum_{i,j} J_{i,j} \sigma_i \sigma_j-\sum_i \mu
\sigma_i\,,
\end{equation}
where $\mu>0$ in the case of micro-canonical ensemble. Here, the interaction strength between occupational numbers $\sigma_i$ is given by $J_{i,j}>0$. The Hamiltonian of Eq. ($\ref{ham}$) is formally analogous to the Ising Hamiltonian in the mean field approximation, with the substitution \cite{b13}
\begin{equation}\label{ljd}
\sigma_i= \frac{S_i + 1}{2}\,,
\end{equation}
where $S_i=\pm1$ represents the Ising spin variable. Once the Hamiltonian is known, it is easy to get the partition function $\mathcal Z$ and to evaluate the thermodynamical variables. The information of network interactions is contained in $J_{i,j}$. Hence, by assuming that the volume and temperature are functions of the redshift, we choose an unitary lattice spacing and a hard sphere interaction, which consists in introducing an excluded volume by prohibiting the multiple occupancy of particles on a given lattice site \cite{eu}. In order to obtain the EoS of our model, we can take into account a finite region of volume $\mathcal{V}$ at a given temperature $\mathcal{T}$, where $\mathcal{N}$ particles are confined to move only on $d$-dimensional discrete lattice points, inside the region under interest \cite{eu2}. The volume is expressed by $\mathcal{V} = h\cdot L^{d-1}$ with $h$ the height and
$L^{d-1}$ the area of the side wall. Thus, the entropy of the system becomes
\begin{equation}\label{ent898989}
\mathcal{S}=k_B\ln{\Omega(\mathcal{N},\mathcal{V})}\,,
\end{equation}
where $\Omega(\mathcal{N},\mathcal{V})$ represents the total permutations of putting $\mathcal{N}$ particles in a volume $\mathcal{V}$. In the absence of excluded volume
interactions, the entropy of the system is factorized
\begin{equation}\label{ent989898989}
\mathcal{S} = k_B \ln\Omega(\mathcal{N},\mathcal{V})=k_B\ln(\omega_i^\mathcal{N})=k_B\mathcal{N}\ln(\omega_i)\,,
\end{equation}
where $\Omega=\omega_i^\mathcal{N}$, and $\omega_i$ represents the single wall probability associated to the $i-$th particle in a unitary volume. In the simplest case of one single particle, the entropy reduces to $k_B\ln\omega$. In \emph{classical} statistical mechanics, the partition function is in general $
{\mathcal Z}=\exp\left(-\frac{1}{k_B \mathcal{T}} {\mathcal H}\right)$ and reduces to a simple form when the Hamiltonian is not a functional. Note that, the contributions to the partition function come from two sources: one is associated to the kinetic energy of the system, while the other one is represented by the total potential energy.

\subsection{The equilibrium configuration}

In the equilibrium configuration, the contribution coming from the kinetic energy is decoupled from the configurational statistics \cite{iniz}. Hence, the equilibrium pressure can be determined by considering only the configurational properties of the system. In the case of a lattice configuration, the free energy is computed exactly, because the hard sphere interaction is formally equivalent to introducing an excluded volume \cite{hjjj}. Hence, by prohibiting multiple particle occupancy on a given lattice site \cite{Binder}, we get
\begin{equation}\label{by}
\Omega(\mathcal{N},\mathcal{V})=\frac{\mathcal{V}!}{\mathcal{N}!(\mathcal{V}-\mathcal{N})!}\,.
\end{equation}
By assuming that the DE density is $\rho\equiv\frac{\mathcal{N}}{\mathcal{V}}\rho_\Lambda$, making use of the Stirling approximation, we infer
\begin{equation}\label{lucky}
\mathcal{S} \approx -k_B \mathcal{V} [\rho \ln\rho + (1-\rho) \ln(1-\rho)]\,,
\end{equation}
where we plugged into Eq. ($\ref{ent989898989}$), the definition of $\Omega(\mathcal{N},\mathcal{V})$, i.e. Eq. ($\ref{by}$). Moreover,  the scaling ruler $\rho_\Lambda$, is associated to the minimal size of lattice sites. For the sake of clearness, $\rho_\Lambda$ determines whether the approximation of lattice fluid holds. Thus, the term $\rho_\Lambda$ defines a \emph{border constant} of the equilibrium configuration, whose physical meaning has nothing to do with  the number density of lattice sites. For our purposes, it is convenient to assume  hereafter $\rho_\Lambda=1$. We replace $\rho_\Lambda$, in the incoming sections. Moreover, the pressure of our fluid, i.e. $\mathcal{P}=\omega\rho$, is given by
\begin{equation}\label{nonsegnata}
\mathcal{P} = - \mathcal{T} \frac{\partial \mathcal{S}}{\partial \mathcal{V}} = -k_B \mathcal{T} \ln (1-\rho)\,,
\end{equation}
where $\omega$ represents the EoS of our model. As shown in \cite{Hong}, the same results could be found in the context of the grand partition ensemble, without losing generality.

The expressions of the pressure and density are interpreted in our picture as sources of DE. They enter the energy momentum tensor, in addition to standard pressureless matter. Thus, for a spatially flat homogeneous and isotropic universe, i.e. $ds^2=dt^2-a(t)^2(dr^2+r^2\sin^{2}\theta d\phi^2+r^2d\theta^2)$, the Friedmann equations are
\begin{eqnarray}\label{ave2}
H^2 &=& {8\pi G\over3}\rho_t\,,\nonumber\\
\,\\
\dot H + H^2&=&-{4\pi G\over3}\left(3\mathcal{P}+\rho_t\right)\,,\nonumber
\end{eqnarray}
where both the total pressure and density, respectively $\mathcal{P}$ and $\rho_t$, are composed by pressureless matter, i.e. $\rho_m$, and by DE density\footnote{Notice that the total pressure $\mathcal{P}$ is the pressure of DE, because standard matter is supposed to be pressureless.}, i.e. $\rho$.

\section{Cosmological consequences of network interacting Ising fluid}

In this section, we are interested in finding the EoS of DE, modeling it as a network interacting fluid through the use of standard thermodynamic recipe. Hence, perfect fluids are modeled by an energy momentum tensor of the form
\begin{equation}\label{mq}
T^{\alpha \beta}=(\rho +\mathcal{P})u^{\alpha} u^{\beta} - \mathcal{P} g^{\alpha \beta}\,,
\end{equation}
from that it naturally follows  the conservation laws for energy  and particle number densities. In particular, by assuming the redshift definition in terms of the cosmic time and $H(z)$, i.e.
\begin{equation}\label{hx}
\frac{dz}{dt} = -(1+z)H(z)\,,
\end{equation}
we get, from Eqs. ($\ref{mq}$) and ($\ref{hx}$), the continuity equation for $\rho$ in terms of $z$
\begin{equation}\label{utyyyy}
\frac{d\rho}{dz}=3\frac{\mathcal{P}+\rho}{1+z}\,.
\end{equation}
From Eq. ($\ref{utyyyy}$), one gets the functional form of $\rho$
\begin{equation}\label{ronzo}
 \rho \propto \exp\left[{3\int{\frac{1 + \omega(z)}{1+z}}dz}\right]\,,
\end{equation}

\noindent and, in addition, it is possible to show that
\begin{equation} \label{temp}
\mathcal{T} \propto  \exp\left[{3\int{\frac{ \omega(z)}{1+z}}dz}\right]\,,
\end{equation}
which represents the DE temperature. It is remarkable to notice that possible temperature measurements at different epochs of the universe could discriminate whether DE dominates over other species or not. Current observations seem to indicate negligible departures from present temperature \cite{alca}. This does not permit us to fix constraints on the DE temperature today, and on its evolution. According to Eqs. ($\ref{ronzo}$) and ($\ref{temp}$), in the small redshift approximation, one recovers the equipartition principle between $\rho$ and $\mathcal{T}$, having $\mathcal{T}\approx\alpha\rho$, with $\alpha$ a constant to be determined. Plugging the temperature definition in terms of $\rho$ into the Eq. ($\ref{nonsegnata}$), we get
\begin{equation}\label{a}
\mathcal{P}=\alpha\rho\log[1-\rho]\,.
\end{equation}
Once the degrees of freedom are fixed, the corresponding cosmological model depends on the today DE value. From Eq. ($\ref{utyyyy}$), we obtain the continuity equation for DE
\begin{eqnarray}\label{EqRhoZ}
\underbrace{\frac{(1+z)}{3}\frac{d\rho}{dz}-\rho}&=&\underbrace{\alpha\rho\log\Big[1-\frac{\rho}{\rho_\Lambda}\Big]}\,,\\
\textsc{Dynamics\,of\,DE}&&\textsc{Source\,of\,DE}\nonumber
\end{eqnarray}
where we restored the definition of $\rho_\Lambda$.  In Eq. (\ref{EqRhoZ}), we split the dynamics of DE in terms of the redshift $z$ (left side), from the DE source (right side), as determined by our model. We evaluate the Hubble rate, by expanding around $\rho_\Lambda$ and we get
\begin{equation}\label{Hz}
H(z)=H_0\sqrt{\Omega_m(1+z)^3+\Big[\frac{\alpha}{\rho_\Lambda}-\frac{\Upsilon}{(1+z)^3}\Big]^{-1}}\,,
\end{equation}
with $\Upsilon$ an integration constant, given by
\begin{eqnarray}\label{b_def}
\Upsilon=\frac{\alpha}{\rho_\Lambda}-\frac{1}{1-\Omega_m}\,.
\end{eqnarray}
It is worth noticing that the error due to the approximations made in Eqs. ($\ref{lucky}$) and ($\ref{Hz}$) is negligibly small. The numerical solution, corresponding to our logarithmic correction of the pressure, well approximates the exact solution, with an error $\leq10\%$, at small $z$. Thus, the EoS of DE can be  rewritten as
\begin{eqnarray}\label{eos}
\omega(z)=-\frac{\alpha}{\rho_\Lambda}\frac{1}{\frac{\alpha}{\rho_\Lambda}-\frac{\Upsilon}{(1+z)^3}}\,,
\end{eqnarray}
whose value today is
\begin{equation}\label{w0}
\omega_0=-\frac{\alpha}{\rho_\Lambda}\left(1 - \Omega_m\right)\,,
\end{equation}
which becomes $\omega_0=-1$ when $\frac{\alpha}{\rho_\Lambda}\rightarrow-\frac{1}{1-\Omega_m}$. From the former relation for $\omega_0$, it is evident that our model reduces to $\Lambda$CDM as a limiting case, showing that the role played by the cosmological constant is actually \emph{mimicked} through the term $\propto \frac{\alpha}{\rho_\Lambda}$. To guarantee that the universe is accelerating today, we should constrain the ratio $\frac{\alpha}{\rho_\Lambda}$. The degeneracy between $\alpha$ and $\rho_\Lambda$ is alleviated by the fact that one needs to evaluate the ratio $\frac{\alpha}{\rho_\Lambda}$, instead of $\alpha$ and $\rho_\Lambda$ separately. Using the Friedmann equations, i.e. Eqs. ($\ref{ave2}$), by keeping in mind the definition of the acceleration parameter:
\begin{equation}\label{quoppo}
q=-1+\frac{(1+z)}{H}\frac{dH}{dz}\,,
\end{equation}
we find for our model
\begin{equation}\label{qu}
q=\frac{ 3\Omega_m(1+z)^6 - 3\Upsilon \frac{\rho_\Lambda^2}{\alpha^2}\omega(z)^2 }{ 2\Omega_m(1+z)^6 + 2\frac{\rho_\Lambda(1+z)^3}{\alpha}\omega(z)} - 1\,.
\end{equation}
The cosmological acceleration starts at the transition redshift $z_{acc}$, i.e.
\begin{equation}\label{q0primario}
z_{acc}\approx\frac{\rho_\Lambda+3(1 - \Omega_m)^2 \alpha}{9(1 - \Omega_m)^2\alpha\Big[1 - (1 - \Omega_m^2) \frac{\alpha}{\rho_\Lambda}\Big] }
\,,
\end{equation}
which approximatively reads $z_{acc}\simeq0.745$, when $\alpha/\rho_\Lambda\approx1.341$, with the indicative value $\Omega_m=0.3$. Equation ($\ref{q0primario}$)  corresponds to the case $q=0$, and leads to the epoch in which DE dominates over standard pressureless matter \cite{unoinpiu}. The acceleration parameter becomes negative, according to current observations, when $z<z_{acc}$. Moreover, the transition redshift, given by Eq. ($\ref{q0primario}$), is in an excellent agreement with the value predicted by $\Lambda$CDM, i.e. $z_{acc,\Lambda}\simeq0.75$. The variation of $q$ with respect to the cosmological redshift $z$, measures the rate of change of the acceleration, i.e. $j(z)$. For our model, it is easy to get
\begin{eqnarray}\label{j}
j(z)=-\rho_\Lambda\frac{\beta_0+ \beta_1 Z(z) +\beta_2 Z(z)^2 + \beta_3 Z(z)^3}{(\rho_\Lambda + Z(z))^2(\rho_\Lambda +\Omega_m Z(z))}\,,
\end{eqnarray}
where for simplicity we defined the following constants
\begin{eqnarray}\label{beta_i}
\beta_0&=&-\rho_\Lambda^2- 9 \alpha^2 (1-\Omega_m)^3+ 9 \alpha (1-\Omega_m)^2 \rho_\Lambda\,,\nonumber\\
\beta_1&=&  (7 - 10 \Omega_m) \rho_\Lambda- 9 \alpha (1-\Omega_m)^2\,,\nonumber\\
\beta_2&=&- 1 - 2 \Omega_m\,,\\
\beta_3&=&- \frac{\Omega_m}{\rho_\Lambda}\,.\nonumber
\end{eqnarray}
and the position
\begin{equation}\label{Zz}
Z(z)=\alpha (1-\Omega_m) \Big[(z+1)^3-1\Big]\,.
\end{equation}
In particular, $j(z)$ is also known in literature as the \emph{jerk parameter} and it is given by the definition $j=\frac{\ddot H}{H^3}-3q-2$ \cite{mine}.
We are interested in determining the jerk parameter at $z=0$. It is expected that $j>0$ today in order to guarantee that for $z>z_{acc}$ the acceleration parameter is positive \cite{visser,mine2}. The jerk parameter today reads
\begin{equation}\label{j_0}
j_0=\frac{\rho_\Lambda^2 - 9 \alpha \rho_\Lambda (1-\Omega_m)^2 + 9 \alpha^2 (1-\Omega_m)^3}{\rho_\Lambda^2}\,,
\end{equation}
and the expected acceleration parameter today reads
\begin{equation}\label{q000}
q_0= \frac{1}{2} \Big[3 \Omega_m - 3 \Upsilon (1-\Omega_m)^2\Big]-1\,.
\end{equation}
By the definitions of $q_0$ and $j_0$, it is possible to find out viable priors for the free parameters, $\alpha$, $\rho_\Lambda$ and $\Omega_m$. In particular, we summarize the priors, that we impose in our computational analysis, in Tab. I, taking into account the results obtained in Eqs. ($\ref{q0primario}$), ($\ref{j_0}$) and ($\ref{q000}$).

\section{Cosmological constraints}

\begin{figure}
\centering
\includegraphics[scale=1]{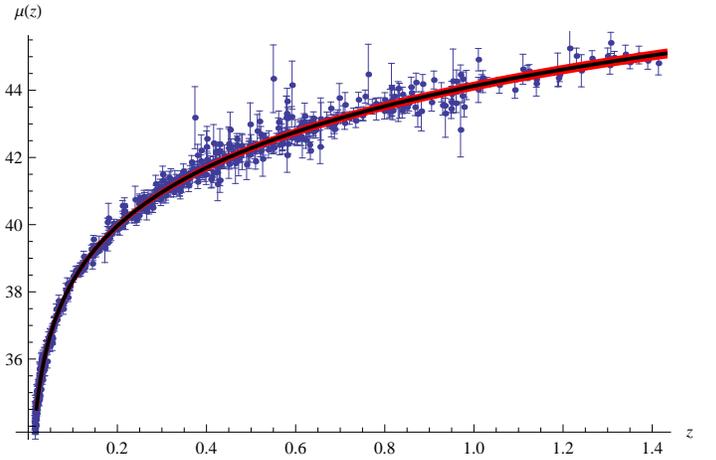}
\caption{Representation of supernova magnitude $\mu(z)$ VS redshift $z$. Data and the best fit, both related to the corresponding uncertainties (red lines). The union 2.1 dataset has been plotted.}
\label{fig:show}
\end{figure}

\begin{figure*}
\begin{center}
\includegraphics[scale=0.8]{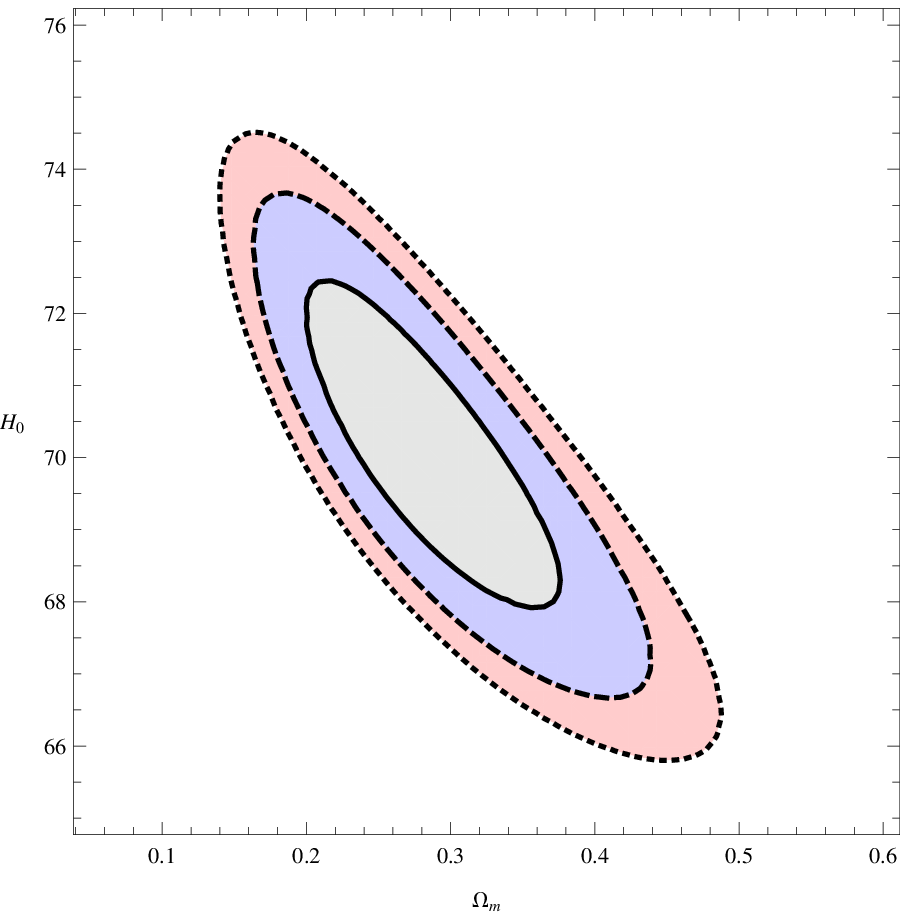}
\includegraphics[scale=0.8]{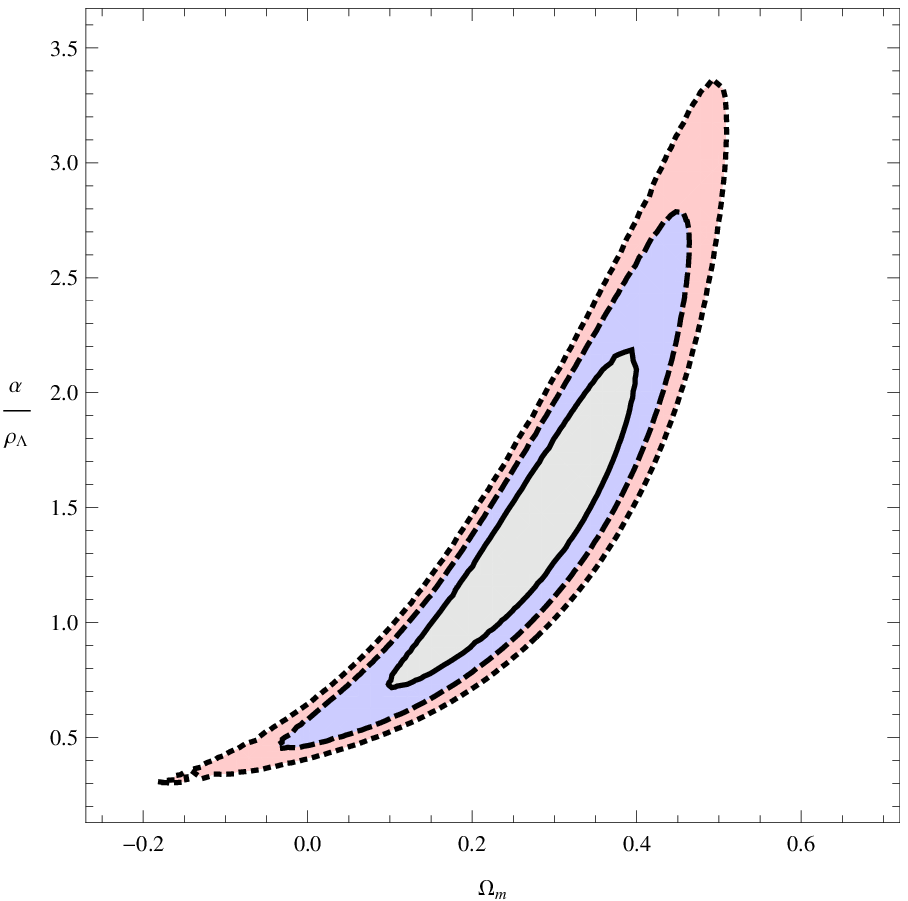}
\includegraphics[scale=0.8]{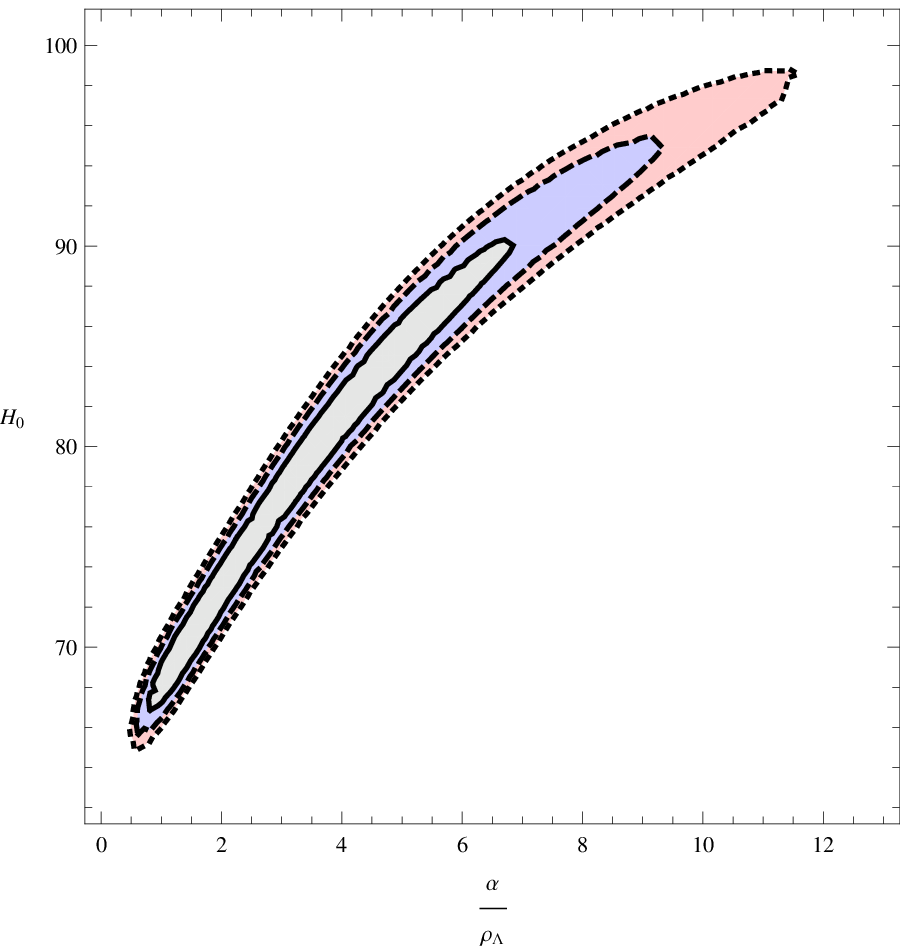}
{\small \caption{1-dimensional marginalized contour plots for
  $H_0$, $\frac{\alpha}{\rho_\Lambda}$ and $\Omega_m$, using $\mathcal A$. We considered the union 2.1 compilation. Here, we report $1\sigma$, $2\sigma$ and $3\sigma$ of confidence levels.}
\label{Fig:EoS}}
\end{center}
\end{figure*}

\begin{figure*}
\begin{center}
\includegraphics[scale=0.8]{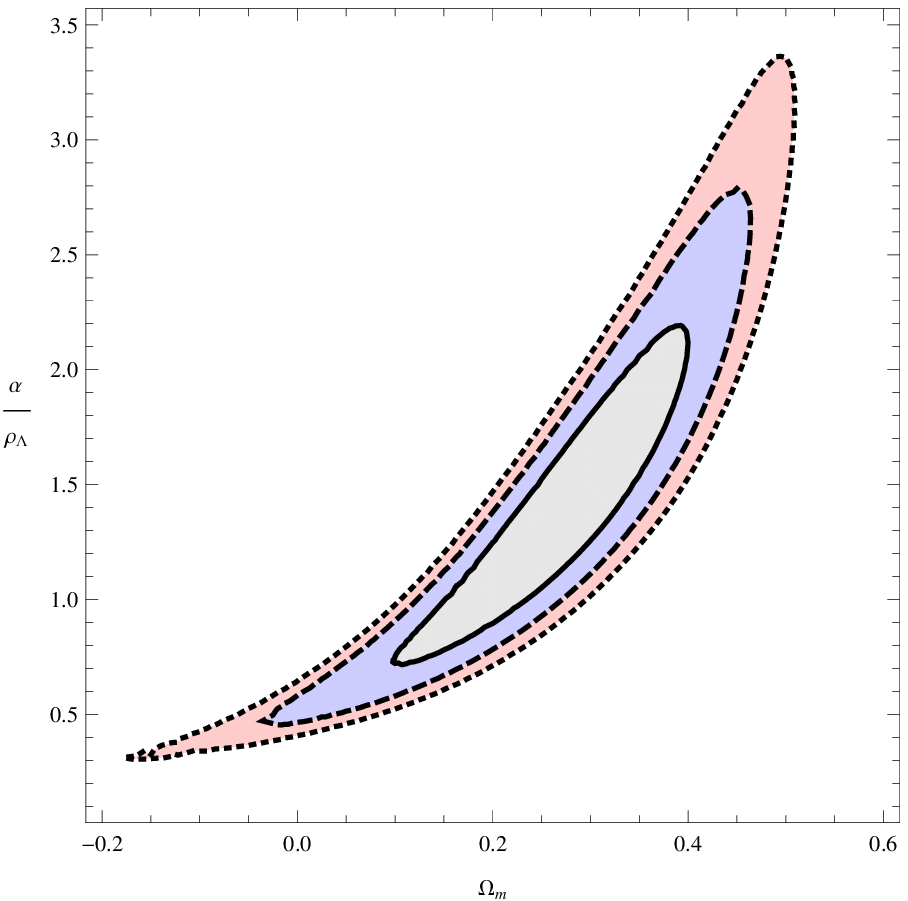}
\includegraphics[scale=0.8]{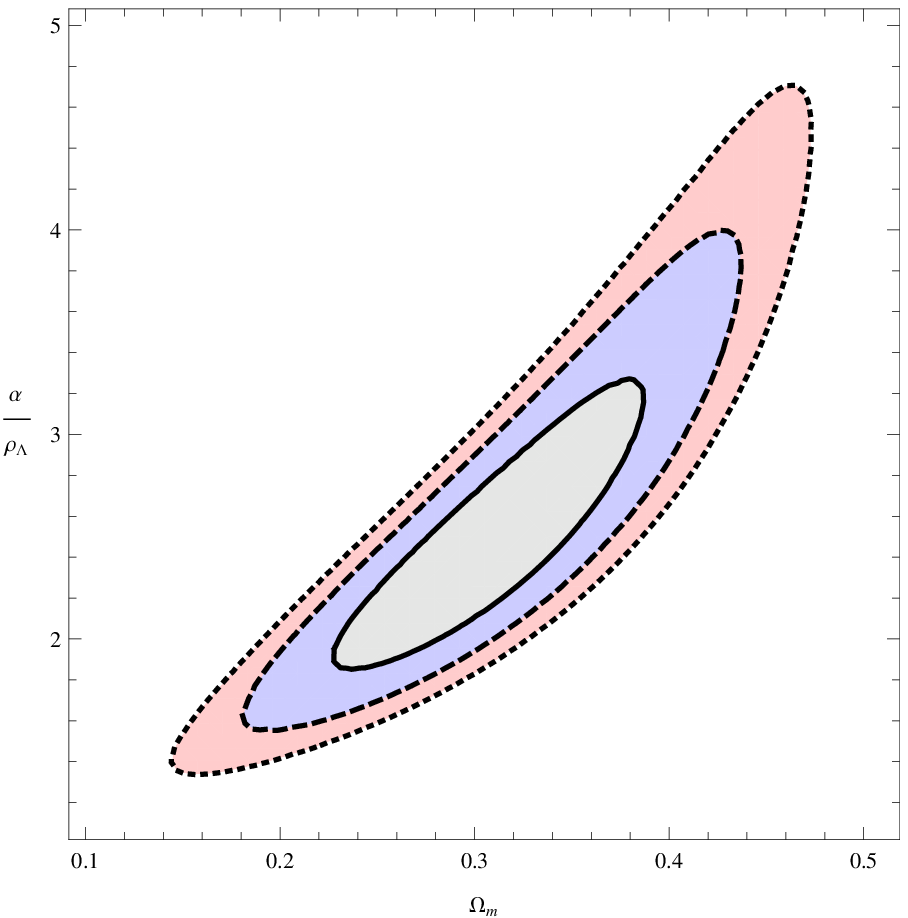}
{\small \caption{1-dimensional marginalized contour plots for
  $H_0$, $\frac{\alpha}{\rho_\Lambda}$ and $\Omega_m$, using set $\mathcal B$. We considered the union 2.1 compilation. Here, we report $1\sigma$, $2\sigma$ and $3\sigma$ of confidence levels. }
\label{Fig:EoS}}
\end{center}
\end{figure*}

\begin{figure*}
\begin{center}
\includegraphics[scale=0.8]{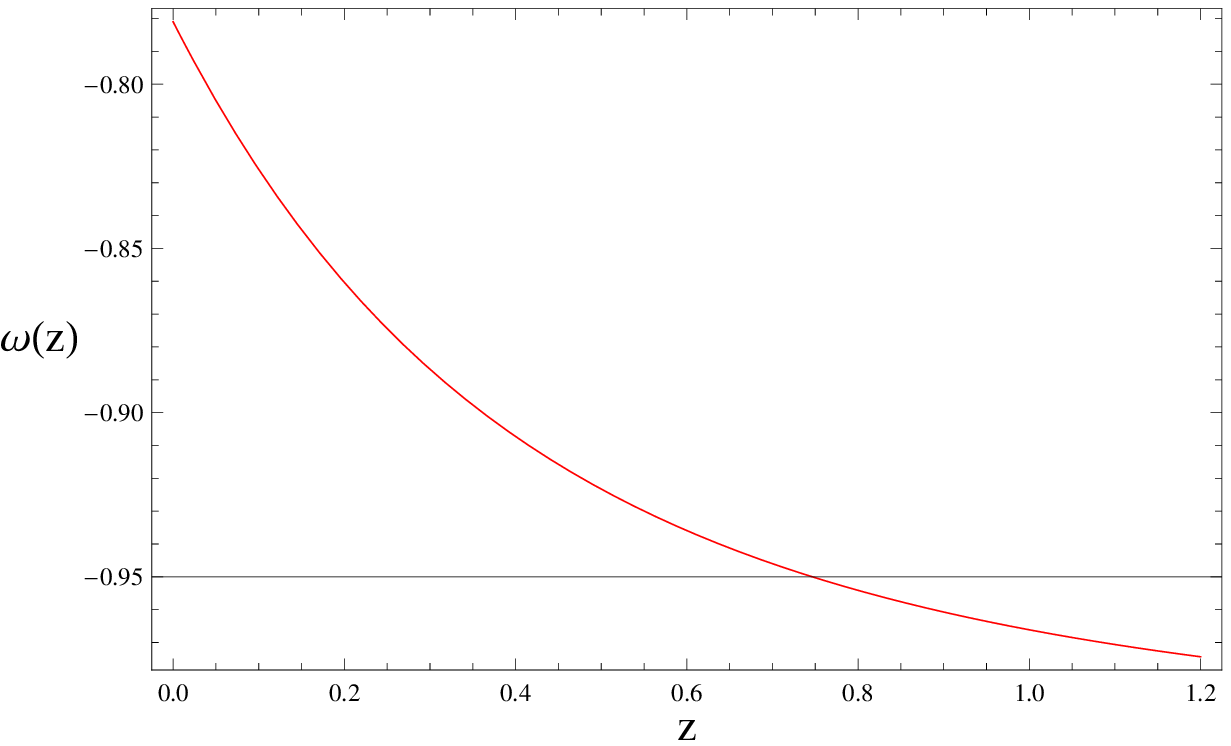}
\includegraphics[scale=0.8]{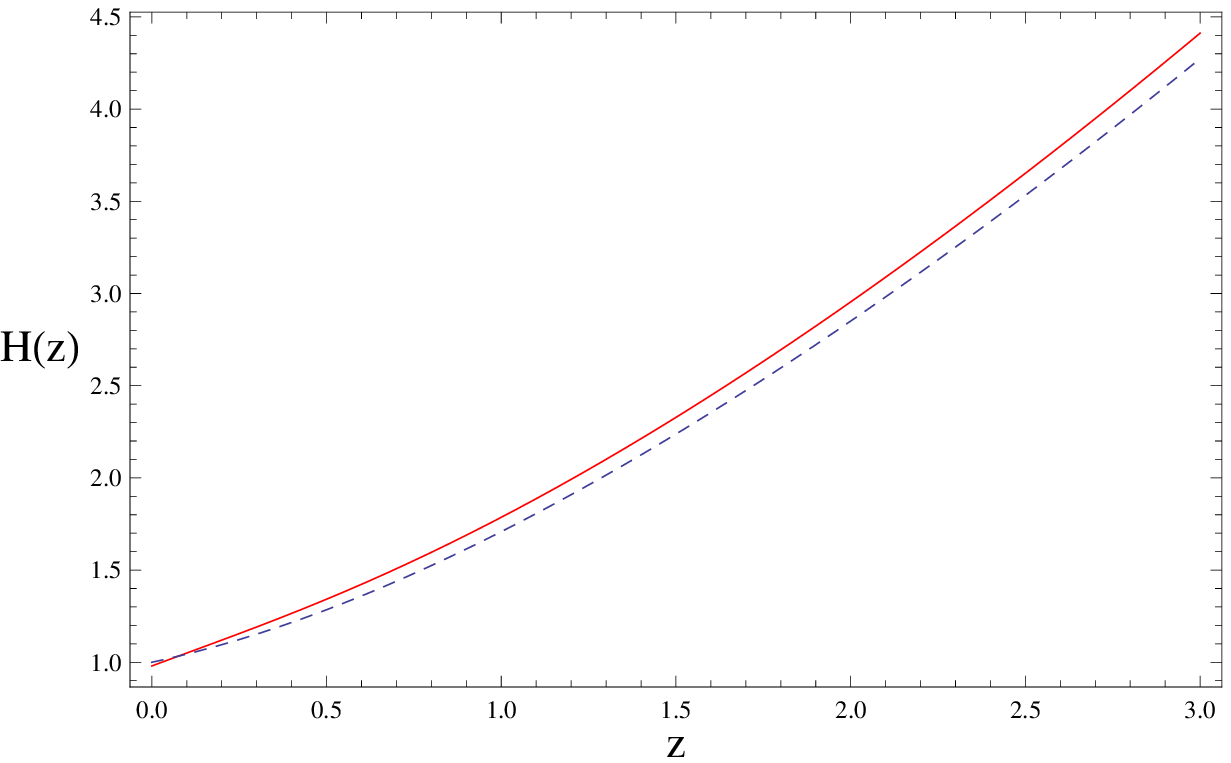}
{\small \caption{We show the behavior of the EoS of our model, in terms of the redshift $z$ (left figure).
Moreover, we plot the Hubble rate (right figure), comparing it with the $\Lambda$CDM model. Respectively red line for our model and dashed line for $\Lambda$CDM. We used the indicative values $\Omega_m=0.290$, $\frac{\alpha}{\rho_\Lambda}=1.1$ for our model, while for $\Lambda$CDM we used the values of WMAP 7-years.}
\label{Fig:EoS}}
\end{center}
\end{figure*}

\begin{figure*}
\begin{center}
\includegraphics[scale=0.8]{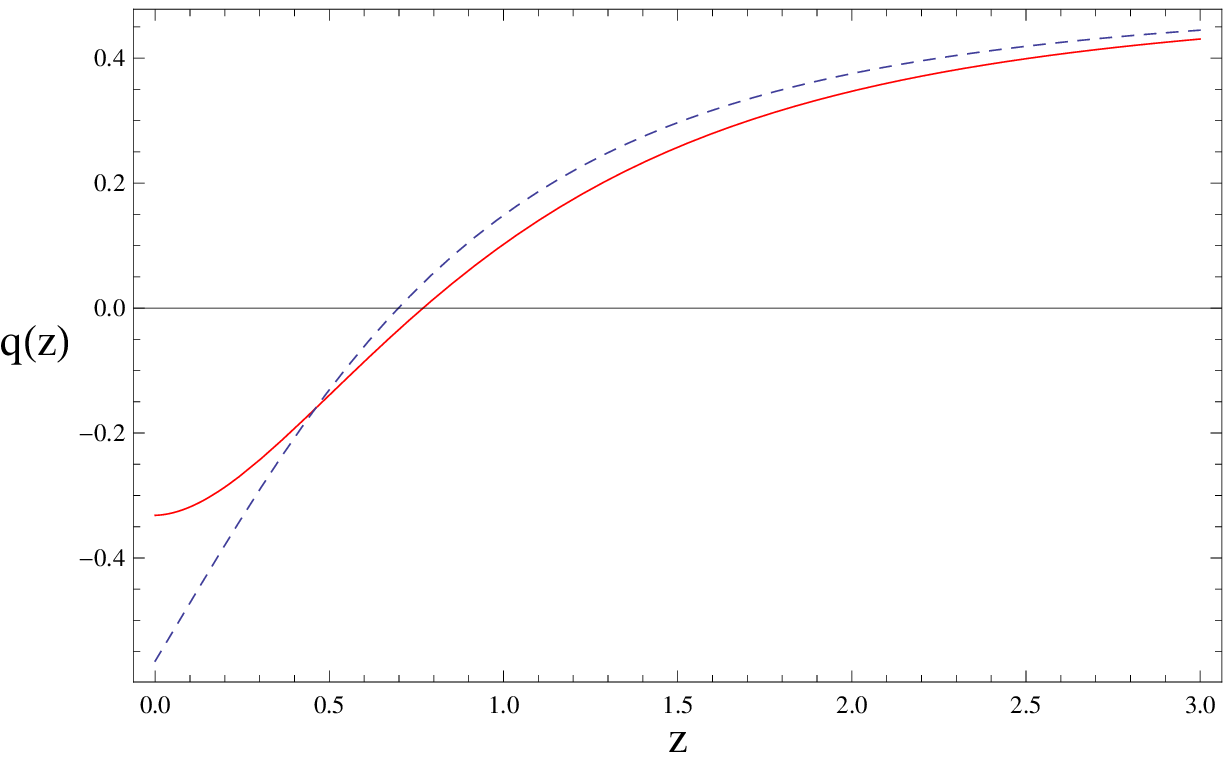}
{\small \caption{Graphic of $q(z)$ for our model (color line) and $\Lambda$CDM (dashed). We notice small differences in the redshift transitions from the acceleration to deceleration phases and good agreement with the numerical value at small redshift with respect to present cosmographic bounds.}
\label{Fig:EoS}}
\end{center}
\end{figure*}

In this section, we constrain the free parameters of our model, through a numerical analysis based on current cosmological data. We rely on three statistical sets of parameters. In doing so, we define three different maximum order of parameters, assuming a hierarchy among the three sets. The sets are summarized as follows

\begin{eqnarray}\label{zucc}
{\mathcal A} &=& \left\{H_0,\,\, \Omega_m,\,\, \frac{\alpha}{\rho_\Lambda}\right\},\;\nonumber\\
{\mathcal B} &=&\left\{\Omega_m,\,\, \frac{\alpha}{\rho_\Lambda}\right\},\;\\
{\mathcal C} &=& \left\{\frac{\alpha}{\rho_\Lambda}\right\}\,.\nonumber
\end{eqnarray}

\noindent The hierarchy of Eqs. ($\ref{zucc}$) predicts a broadening of the sampled distributions, adding the cosmological coefficients to  constrain.  In the Gaussian regime, the error propagation becomes higher as hierarchy,  between parameters,  increases. To alleviate the error propagation and possible systematics, we consider the numerical priors reported in Tab. I.

\subsection{Priors on the cosmological parameters and the initial condition on $H_0$}

Viable cosmological priors are need in order to alleviate the so called degeneracy between cosmological distances. In fact, the luminosity distance by itself is not enough to separately constrain all the cosmological densities. It follows that the EoS cannot be constrained with arbitrary accuracy with SNeIa data only. By fixing viable priors, one needs to reduce the total phase space. This permits us to complement distance measurements by different constraints. In other words, once the priors have been determined, it is possible to infer cosmological bounds from one class of measurements. Moreover, the simple choice of using different distances does not guarantee \emph{a priori} that the physical region for constraining the free parameters is actually reduced. For those reasons, we rely on three combined tests, performed by using SNeIa, BAO measurements and the CMB surveys. Our choice determines a combination of cosmological and geometrical procedures, which allows to circumscribe the phase space with higher precision. In addition, spatial geometry is also set to be geometrically flat, while the initial condition on $H_0$ is determined as follows: $1)$ first, we consider $H_0$ free to vary, $2)$ second, we constrained it by WMAP 7-years results, $3)$ finally, we impose its value, by assuming the Hubble space telescope (HST) measure. Additional cosmological and geometrical priors, adopted throughout our numerical analysis, are summarized in Tab. I, as already underlined.

\newlength{\mywidth}
\setlength{\mywidth}{0.4\textwidth}
\begin{table}
\begin{center}
\begin{tabular}{c}
\begin{tabular*}{\mywidth}{c}
\hline
Flat priors\\ \hline \hline
\begin{tabular}{rcl}
$0.5\quad <$ &$h$ & $< \quad 0.8$\\
$0.009 \quad <$&$\Omega_{\rm b}h^2$ & $< \quad 0.08$ \\
$0.05 \quad <$&$\Omega_{\rm dm}h^2$ & $< \quad 0.26$ \\
$0.01\,\rho_\Lambda \quad <$&$\alpha$ & $< \quad 100\,\rho_\Lambda$ \\
\end{tabular}\\\hline\\  \hline
\hline
Additional constraints\\ \hline \hline
\begin{tabular}{rl}
$\Omega_{k}$&$ =\quad0$\\
$\Omega_{m}$ & $< \quad 0.33$\\
\end{tabular}\\ \hline
\end{tabular*}
\end{tabular}
\end{center}
\caption{Priors imposed on the free parameters, involved in the the Bayesian analyses. We considered the additional geometrical assumption of a spatially flat universe.}\label{tab:priors}
\end{table}

\subsection{Cosmological datasets}
\begin{table*}
\caption{
Best fits of the free parameters of our model, tested by SNeIa. The quoted errors show the 68.3\%, 95.4\% and 99.7\% confidence level uncertainties.}

\begin{tabular}{c|c|c|c|c|c} 
\hline\hline\hline 
{\small Parameter} & Set ${\mathcal A}$ & Set ${\mathcal B}$  & Set ${\mathcal B}$  & Set ${\mathcal C}$ & Set ${\mathcal C}$ \\
 & $\chi^2_{min}= 0.9727 $  & $\chi^2_{min}= 0.9727 $ &$\chi^2_{min}=0.9756 $ & $\chi^2_{min}=0.9728$ & $\chi^2_{min}=0.9837$\\
\hline
${\mathcal H}_0$ & {\small $70.2$}{\tiny${}^{+1.3+2.0+2.5}_{-1.3-2.0-2.4}$}  & ${\mathcal H}_0\equiv70.2$ &  ${\mathcal H}_0\equiv74.2$  & ${\mathcal H}_0\equiv70.2$  & ${\mathcal H}_0\equiv74.2$ \\
\hline
$\Omega_m$ & {\small $0.278$}{\tiny${}^{+0.052+0.083+0.10}_{-0.048-0.072-0.088}$} & {\small $0.278$}{\tiny${}^{+0.052+0.083+0.10}_{-0.048-0.072-0.088}$} & {\small $0.305$}{\tiny${}^{+0.038+0.060+0.075}_{-0.036-0.055-0.068}$} & $\Omega_m\equiv0.274$ & $\Omega_m\equiv0.274$ \\
\hline
$\frac{\alpha}{\rho_\Lambda}$  & {\small $1.42$}{\tiny${}^{+0.27+0.42+0.52}_{-0.26-0.39-0.48}$} & {\small $1.42$}{\tiny${}^{+0.27+0.41+0.52}_{-0.26-0.39-0.48}$} & {\small $2.44$}{\tiny${}^{+0.32+0.51+0.58}_{-0.31-0.47-0.64}$} & {\small $1.40$}{\tiny${}_{-0.25-0.39-0.47}^{+0.26+0.41+0.51}$}    & {\small $2.22$}{\tiny${}_{-0.26-0.42-0.52}^{+0.26+0.44+0.56}$}  \\

\hline\hline\hline
\end{tabular}

{\footnotesize
Notes.
${\mathcal H}_0$ is given in Km/s/Mpc.}
\label{table:summary} 
\end{table*}

\begin{table*}
\caption{
Best fits of the free parameters with BAO. Note that with BAO it is not possible to estimate ${\mathcal H}_0$. The quoted errors show the 68.3\%, 95.4\% and 99.7\% confidence level uncertainties.}

\begin{tabular}{c|c|c} 
\hline\hline\hline 
$\qquad${\small Parameter}$\qquad$ & $\qquad$ Set ${\mathcal A}$ $\qquad$& $\qquad$ Set ${\mathcal C}$ \\
 & \, & \,  \\[0.2ex]
\hline
$\Omega_m$ & {\small $0.285$}{\tiny${}^{+0.030+0.043+0.053}_{-0.027-0.038-0.046}$} & $\Omega_m\equiv0.274$ \\[0.8ex]
\hline
$\frac{\alpha}{\rho_\Lambda}$      & {\small $1.33$}{\tiny${}^{+0.36+0.59+0.74}_{-0.41-0.50-0.59}$} & {\small $1.48$}{\tiny${}_{-0.37-0.51-0.62}^{+0.42+0.61+0.76}$} \\

\hline\hline\hline
\end{tabular}

\label{table:summary} 
\end{table*}

\begin{table*}
\caption{
Best fits of the free parameters with CMB. Note that with CMB it is not possible to estimate ${\mathcal H}_0$. The quoted errors show the 68.3\%, 95.4\% and 99.7\% confidence level uncertainties.}

\begin{tabular}{c|c|c} 
\hline\hline\hline 
$\qquad${\small Parameter}$\qquad$   &  $\qquad$ Set ${\mathcal A}$ $\qquad$& $\qquad$ Set ${\mathcal C}$ $\qquad$ \\
 & \, & \,  \\[0.2ex]
\hline
$\Omega_m$ & {\small $0.281$}{\tiny${}^{+0.034+0.050+0.062}_{-0.030-0.041-0.050}$}   &  $\Omega_m\equiv0.274$ \\
\hline
$\frac{\alpha}{\rho_\Lambda}$ & {\small $1.17$}{\tiny${}^{+0.30+0.45+0.57}_{-0.23-0.31-0.37}$}   & {\small $1.24$}{\tiny${}_{-0.24-0.33-0.39}^{+0.31+0.47+0.60}$}  \\

\hline\hline\hline
\end{tabular}

\label{table:summary} 
\end{table*}

\begin{table*}
\caption{
Best fits of the parameters for the three considered models with SNeIa, BAO and CMB. The quoted errors show the 68.3\%, 95.4\% and 99.7\% confidence level uncertainties.}

\begin{tabular}{c|c|c|c|c|c} 
\hline\hline\hline 
{\small Parameter} & Set ${\mathcal A}$ & Set ${\mathcal B}$  & Set ${\mathcal B}$  & Set ${\mathcal C}$ & Set ${\mathcal C}$ \\
 & $\chi^2_{min}= 0.9727 $  & $\chi^2_{min}= 0.9727 $ &$\chi^2_{min}=0.9871 $ & $\chi^2_{min}=0.9729$ & $\chi^2_{min}=0.9885$\\
\hline
${\mathcal H}_0$ & {\small $70.2$}{\tiny${}^{+1.3+2.0+2.5}_{-1.3-2.0-2.4}$}  & ${\mathcal H}_0\equiv70.2$ &  ${\mathcal H}_0\equiv74.2$  & ${\mathcal H}_0\equiv70.3$  & ${\mathcal H}_0\equiv74.2$ \\
\hline
$\Omega_m$ & {\small $0.278$}{\tiny${}^{+0.049+0.078+0.099}_{-0.045-0.068-0.083}$} & {\small $0.278$}{\tiny${}^{+0.049+0.078+0.99}_{-0.045-0.068-0.083}$} & {\small $0.285$}{\tiny${}^{+0.032+0.056+0.072}_{-0.031-0.051-0.064}$} & $\Omega_m\equiv0.274$ & $\Omega_m\equiv0.274$ \\
\hline
$\frac{\alpha}{\rho_\Lambda}$ & {\small $1.42$}{\tiny${}^{+0.26+0.41+0.51}_{-0.25-0.39-0.48}$} & {\small $1.42$}{\tiny${}_{-0.25-0.39-0.48}^{+0.26+0.41+0.51}$} & {\small $2.28$}{\tiny${}_{-0.25-0.42-0.53}^{+0.25+0.44+0.57}$} & {\small $1.40$}{\tiny${}_{-0.25-0.38-0.47}^{+0.26+0.40+0.51}$} & {\small $2.21$}{\tiny${}_{-0.23-0.40-0.51}^{+0.24+0.43+0.55}$}  \\[0.8ex]

\hline\hline\hline
\end{tabular}

{\footnotesize
Notes.
${\mathcal H}_0$ is given in Km/s/Mpc.}
\label{table:summary} 
\end{table*}

In our numerical analysis, we consider three different datasets. In particular, we take into account the union 2.1 compilation, the BAO measure and the CMB surveys, with the constraint on $H_0$ given by the measurement of the WMAP 7-years and HST respectively. In the union 2.1 compilation of the supernova cosmology project \cite{Suzuki:2011hu}, the covariance matrix has been evaluated with and without systematics. The compilation includes previous surveys, i.e. union 2 \cite{Amanullah:2010vv} and union 1 \cite{Kowalski:2008ez}. The supernova measurements may be represented in the plane modulus-redshift, i.e. $\mu-z$. They consist of 580 measurements of $\mu$  and $z$, spanning in the redshift range $ 0.015 < z < 1.414$. The error over $z$ is assumed to be negligibly small, while to each supernova is associated a corresponding error on $\mu$. This turns out to be important since supernovas are standard indicators and  represent the primary distance indicators. The relevance of supernova measurements is related to the fact that their rest frame wavelength region spans from 4000 to 6800 ${\AA}$ for all the transient events which characterize supernovas. By assuming that the same rest frame wavelengths are measured at all $z$, one can compare the supernova brightness as independent as possible from a supernova model. Thence, to fix cosmological constraints, we make use of a Bayesian method in which
the best fits of parameters are inferred by maximizing the following likelihood function
\begin{equation}\label{funzionea}
{\mathcal L} \propto \exp(-\chi^2/2)\,,
\end{equation}
where $\chi^2$ is the  ({\it pseudo}){\it chi-squared} function. To obtain the corresponding posterior distributions, we consider uniform priors in the interval ranges of Tab. I. It is easy to show that the luminosity distance is
\begin{equation}
d_L(z) =  (1+z) \int_0^z \frac{d\xi}{\mathcal{E}(\xi)}\,,
\end{equation}
where $\mathcal{E}\equiv\frac{H}{H_0}$, and by defining the distance modulus $\mu$ for each supernova
\begin{equation}
\mu = 25 + 5 \log_{10} \frac{d_L}{Mpc}\,,
\end{equation}
together with the corresponding 1-$\sigma_i$ error, we can rewrite the $\chi^2$ parameter of Eq. (\ref{funzionea}) as
\begin{equation}
\chi^{2}_{SN} =
\sum_{i}\frac{(\mu_{i}^{\mathrm{theor}}-\mu_{i}^{\mathrm{obs}})^{2}}
{\sigma_{i}^{2}}\,.
\end{equation}
A simple test with SNeIa minimizes the term $\chi_{SN}^{2}$, which corresponds to maximize Eq. ($\ref{funzionea}$).

The large scale galaxy clustering observations provide the signatures of the BAO \cite{Percival:2009xn}. The theoretical background employs that the universe consisted of a hot plasma of photons, electrons, protons, baryons and other light nuclei, at a certain epoch of its evolution. As a consequence, the Thompson scattering between photons and electrons leads to oscillations in the hot plasma. As the universe becomes neutral, it is possible to consider the initial perturbation patterns which are imprinted on the matter distribution. Hence, by observing the spectrum of galaxy correlations today, it is possible to focus on observations of BAO. The corresponding measurement is represented by a sound horizon length, whose physical meaning deals with the distance traveled by an acoustic wave by the time of plasma recombination. The BAO peak is considered a standard cosmological ruler, because such a peak is independent of the choice of a particular cosmological model. We use the peak measurement of luminous red galaxies,  denoted by $A_{BAO}$. It reads
\begin{eqnarray}
A_{BAO}=&&\left(\Omega_b+\Omega_{dm}\right)^{\frac{1}{2}}  \Big[\frac{1}{\mathcal{E}(z_{BAO})}\Big]^{\frac{1}{3}}\times\nonumber\\
&&\times\left[ \frac{1}{z_{BAO}}\int_0^{z_{BAO}}
\frac{1}{\mathcal{E}(\xi)}d\xi\right]^{\frac{2}{3}}\,,
\end{eqnarray}
with $\Omega_b$ and $\Omega_{dm}$ respectively the baryonic and dark matter densities and $z_{BAO}=0.35$. In addition, the observed $A_{BAO}$ is estimated to
be
\begin{equation}\label{ax}
A_{BAO,obs} = 0.469 \left(\frac{0.95}{0.98}\right)^{-0.35}=0.477\,,
\end{equation}
with an error $\sigma_A = 0.017$. In the case of the BAO measurement,
we minimize the chi square
\begin{equation}
\chi^{2}_{BAO}=\left(\frac{A-A_{obs}}{\sigma_A}\right)^2\,.
\end{equation}
The third test is represented by CMB observations. This kind of measurements has recently reached much interest, as a nearly isotropic background was discovered in 1965. Measures of the WMAP satellite found new data, able to alleviate the cosmological degeneracy between models. The underlying philosophy relies on CMB radiation, which can be directly detected by keeping in mind that it is influenced by two cosmological epoches, i.e. the  last scattering era and present time \cite{salone,rubi}. For the CMB test, we define the so-called CMB shift parameter $R_{CMB}$. Its standard definition reads
\begin{equation}\label{cmbsod}
R_{CMB}=\left(\Omega_b+\Omega_{dm}\right)^{\frac{1}{2}} \int^{z_{rec}}_{0}
\frac{d\xi}{\mathcal{E}(\xi)}\,,
\end{equation}
with $z_{rec}\approx1000$. Equation ($\ref{cmbsod}$) presents several theoretical shortcomings \cite{Melch}. It is interesting to replace such a relation with $R_{CMB} \equiv 2 \frac{l_1}{l_1^{'}}$, where $l_1$ is the position of the first peak on the CMB TT power spectrum of the model under consideration. Moreover, $l_1^{'}$ is the first peak in a flat homogeneous and isotropic universe with $\Omega_b=1-\Omega_{dm}$. Hereafter, $l_1$ is written as
\begin{equation}
l_1 = D_A(z_{rec})s(z_{rec})^{-1}\,,
\end{equation}
where $D_A(z_{rec})$ is called co-moving angular distance, i.e.
\begin{equation}
D_A(z_{rec}) = \int^{z_{rec}}_{0} (1+\xi) d\xi\,,
\end{equation}
with $s(z_{rec})$ representing the sound horizon at recombination
\begin{equation}\label{shor}
s(z_{rec}) = \frac{1}{H_0} \int_{z_{rec}}^{\infty}\mathfrak{v}(\xi)\mathcal{E}(\xi)^{-1}d\xi\,.
\end{equation}
We defined  $\mathfrak{v}(\xi)$ as the sound speed of the photon-to-baryon fluid. Its expression reads $\mathfrak{v}= \left(3+4\frac{\rho_b}{\rho_{\gamma}}\right)^{-0.5}$.

\begin{table*}
\small
\caption{BIC and AIC analysis, performed by assuming model $\mathcal{A}$ with SNeIa data. \label{tbl-2}}
\begin{tabular}{@{}crrrrrrrrrrr@{}}
\hline
Model & Num. of Par. (k) & Parameters & $\chi^2_{min}$ & $\Delta BIC$ & $\Delta AIC$ \\
\hline
$\Lambda$CDM &1 & $\Omega_m$ &0.9727 & 0 & 0 \\
Ising Fluid &2 & $\Omega_m$, $\frac{\alpha}{\rho_\Lambda}$ &0.9727 & 6.363 & 2 \\
CPL & 3 & $\Omega_m$, $\omega_0$, $\omega_a$  & 0.9879  & 12.741 & 4.015 \\
\hline
\end{tabular}
\end{table*}

The importance of using CMB leads to its complementary with respect to SNeIa and BAO measurements. This comes from the fact that SNeIa and BAO are confined in low redshift regimes, constrained to $z<2$, while for CMB, the recombination redshift is higher of three orders of magnitude  \cite{Wang:2007mza}.
According to the CMB measurement, we minimize the following
\begin{equation}
\chi^{2}_{CMB}=\left(\frac{{\mathcal R}-{\mathcal R}_{obs}}{\sigma_R}\right)^2\,.
\end{equation}
Since the three different sets of observations are not correlated between them, the total $\chi^2$ reads
\begin{equation}
\chi^2 = \sum_{i=1}^{3}\chi^{2}_{i}=\chi^2_{SN} + \chi^2_{BAO}+\chi^2_{CMB}\,.
\end{equation}

\noindent From one hand, the SNeIa test needs to fix $H_0$, as initial condition, in addition to the free parameters of our model, i.e. $\alpha$ and $\rho_\Lambda$. It is prominent to assume gaussian priors on the Hubble constant, in order to fulfill the initial condition on $H_0$. In doing so, we consider the WMAP 7-years measurements, which suggests ${\mathcal H}_0 = 70.2 \pm 3.6 \, km/s/Mpc$ and $H_0 = 74.2 \pm 3.6 \, km/s/Mpc$ \cite{tabu}, as measured by the HST. Following these constraints, and those of Tab. I,  we perform three separate tests with SNeIa, BAO and finally CMB, by combining them together. On the other hand, the cosmological tests BAO and CMB do not directly depend on $H_0$. It follows that the unique bound on $H_0$ should be imposed when we perform the supernova test only.

\section{Model comparison through AIC and BIC selection criteria}

As already stressed in Sec. III, the degeneracy problem, between cosmological models, plagues standard techniques of numerical analyses. The task of discriminating which models better fit current data is actually a thorny issue of statistics. The main disadvantage lies on assuming, in a numerical computation, a particular form of $H(z)$ for the cosmological model. In other words, each cosmological test assumes \emph{a priori} that the cosmological model under exam is statistically favored. This is a consequence of the $\chi^2$ analysis, which is able to constrain the free parameters of a given model, although it does not provide any information about the validity of the model itself. To alleviate this problem, one could use model independent procedures of statistical analysis.  Among various possibilities, there exist in the literature statistical methods, able to understand which model is really favorite, than others, once the same data survey is used. It is easy to show that comparing different chi squares among alternative models, with the same numbers of free parameters, it is possible to check which model is statistically favored than others. In lieu of limiting our attention to simply fit our model, as in Sec. IV, we compare our approach with two relevant paradigms, i.e. a variable quintessence model: the Chevallier, Polarsky, Linder (CPL) parametrization \cite{cpl1,cpl2}) and the standard $\Lambda$CDM model. Naively, we can expect that  a combination of lowest chi squares and fewest numbers of parameters provides the model which better reproduces cosmological data. We summarize below the $\Lambda$CDM and CPL Hubble rates respectively, i.e.
\begin{eqnarray}\label{ciao}
\mathcal{E}_1&=&\Big[\Omega_m(1+z)^3+1-\Omega_m\Big]^{\frac{1}{2}}\,,\nonumber\\
\,\\
\mathcal{E}_2&=&\Big[\Omega_m(1+z)^3+\left(1-\Omega_m\right)\mathrm{f}(z)\Big]^{\frac{1}{2}}\,,\nonumber
\end{eqnarray}
with $\mathrm{f}(z)=(1+z)^{3(1+\omega_0+\omega_a)}\exp\left(-3\omega_a\frac{ z}{1+z}\right)$. In particular, $\omega=-1$ and $\omega=\omega_0+\omega_a\left(\frac{z}{1+z}\right)$, respectively for $\Lambda$CDM and CPL models. In what follows, we describe the statistical methods that we are going to use, in order to make a comparison between our model and the above cited $\Lambda$CDM and CPL paradigms.

\subsection{Selection criteria}

In this subsection, we propose two statistical methods which follow the guidelines given in Sec. V. They are the so-called AIC \cite{AIC1,AIC2,AIC3,AIC4} and BIC \cite{BIC} selection criteria. The first one has reached a widely accepted consensus, becoming a common diagnostic tool \cite{trs,do,qua,quu}. It has been largely used for regression models \cite{tro}, since its first applications \cite{AIC1,AIC2,AIC3,AIC4}. The basic demands of AIC and BIC consist in postulating two statistical distributions, i.e. $f(x)$ and $g(x|\theta)$. The first distribution $f(x)$ is imposed to be \emph{the exact} reconstruction of a particular model, while the second distribution, i.e. $g(x|\theta)$, is thought to approximate $f(x)$, by bounding $N$ parameters, that are included within the vector $\theta$. Once $f(x)$ and $g(x|\theta)$ are defined by numerical procedures, the set of parameters $\theta$ is estimated, minimizing the departures between $f(x)$ and $g(x,\theta_{min})$.

Obviously, the function $f(x)$ is not known a priori. Hence, both the AIC and BIC criteria are meaningless if evaluated for a single model. In other words, once the minima of AIC and BIC are determined, i.e. $AIC_{min}$ and $BIC_{min}$ respectively, we should investigate the differences $\Delta AIC \equiv AIC - AIC_{min}$ and $\Delta BIC \equiv BIC - BIC_{min}$. A general form for the $AIC$ function is
\begin{equation}
{\rm AIC}=-2\ln{\mathcal L}_{max}+2k\,,
\end{equation}
and by following \cite{BIC}, we define the BIC function as
\begin{equation}
{\rm BIC}=-2\ln{\mathcal L}_{max}+k\ln N\,,
\end{equation}
where for both the methods AIC and BIC, ${\mathcal L}_{max}$ is the maximum likelihood function, corresponding to the minimum of $\chi^2$. Moreover, $k$ is the number of parameters to be estimated, and $N$ is the number of data points used to perform the cosmological fits. To Gaussian distributed  errors corresponds $\chi_{min}^2=-2\ln{\cal L}_{max}$ and we can simplify $\Delta{\rm AIC}=\Delta\chi_{min}^2+2\Delta k$ and $\Delta{\rm BIC}=\Delta\chi_{min}^2+\Delta k\ln N$.

In Tab. III, we report the numerical results of using the AIC and BIC criteria. In particular, we find for $\mathcal{E}_1$ and $\mathcal{E}_2$ respectively: $\Omega_m=0.279^{+0.016}_{-0.016}$, $H_0=69.966^{+1.291}_{-1.270}$, $\chi_{SN,min}=0.9727$, $\Omega_m=0.282^{+0.023}_{-0.018}$, $H_0=70.120^{+1.414}_{-1.362}$, $\omega_0=-0.95^{+0.49}_{-0.46}$, $\omega_a=1.52^{+1.64}_{-1.59}$ and $\chi_{SN,min}=0.9879$. We find that our model fairly good adapts to cosmological data, being disfavored with respect to the standard $\Lambda$CDM picture, and  behaving better than CPL. The statistical success of $\Lambda$CDM is clearly due to the smallest number of parameters involved into calculations. However, our paradigm shows small departures from $\Lambda$CDM, behaving much better than an evolving quintessence model. This can be interpreted as a possible indication that our network interacting Ising fluid may be seen as a relevant alternative to $\Lambda$CDM.

\section{Final remarks}

In this work, we investigated the possibility to model DE through an Ising fluid on a lattice with network interactions. A negative pressure, associated to DE and compatible with present observations, emerged by considering the barotropic EoS of our model in the equilibrium configuration. We inferred theoretical constraints in a fairly good agreement with current observations.  We demonstrated that, at low redshift, it is possible to obtain a viable Hubble rate which well mimics the DE effects, reducing to $\Lambda$CDM at the zero order expansion, in terms of the DE density. The corresponding acceleration parameter and its variation, namely the jerk parameter, provided interesting results which properly fitted with modern observations. In particular, the acceleration starts at a redshift which is excellently close to the one predicted by $\Lambda$CDM. In addition, the present value of $j(z)$ confirmed that the acceleration parameter has changed its sign, as the universe expands. From such considerations, it followed that one of the main advantages of our model relied on interpreting the DE nature as an emergent Ising fluid with network interactions. In addition, we fixed constraints on the free parameters of our model, by employing three cosmological datasets, i.e. the SNeIa, BAO and CMB surveys. In particular, we first used SNeIa, BAO and CMB separately and then we combined SNeIa with BAO and CMB, constraining the free parameters of our model in tighter intervals, though viable geometrical and cosmological priors. In doing so, we chose for the computational analysis three cosmological sets of observables, ordered in a hierarchial way, evaluating the corresponding errors up to 3$\sigma$ of confidence level. The results confirmed that our model could be viewed as a viable alternative to $\Lambda$CDM, in order to describe the DE effects at present time. To this end, we showed that, through the use of the AIC and BIC selection criteria, our model provided small departures than $\Lambda$CDM, behaving smoother than the so called CPL parametrization. Future efforts could be devoted to apply the network interacting Ising model to other stages of the universe evolution, wondering whether the model could be modeled for different epochs of the universe evolution.

\end{document}